\documentclass[aps,prl,twocolumn,superscriptaddress,amsfont,graphicx,nofootinbib,preprintnumbers]{revtex4}%
\usepackage{color,graphicx,epsfig}
\usepackage{ifpdf}
\usepackage{amsmath}
\usepackage{bm}
\usepackage{color}
\usepackage[english]{babel}
\usepackage{graphicx}%
\usepackage{amsfonts}%
\usepackage{amssymb}
\usepackage{braket}
\usepackage{hyperref}
\usepackage{enumerate}

\definecolor{nicered}{rgb}{0.7,0.1,0.1}
\definecolor{nicegreen}{rgb}{0.1,0.5,0.1}
\hypersetup{colorlinks,citecolor= nicegreen,linkcolor= nicered}

\usepackage{graphicx}
\usepackage{dcolumn}
\usepackage{bm}
\usepackage{slashed}

\begin{document}

\title{Single top squark production as a probe of natural supersymmetry at the LHC}

\author{Ken-ichi Hikasa}
\email[Electronic address:]{hikasa@phys.tohoku.ac.jp}
\affiliation{Department of Physics, Tohoku University, Sendai 980-8578, Japan}

\author{Jinmian Li}
\email[Electronic address:]{jinmian.li@adelaide.edu.au}
\affiliation{ARC Centre of Excellence for Particle Physics at the Terascale, Department of Physics, University of Adelaide, Adelaide, SA 5005, Australia}

\author{Lei Wu}
\email[Electronic address:]{leiwu@physics.usyd.edu.au}
\affiliation{ARC Centre of Excellence for Particle Physics at the Terascale, School of Physics, The University of Sydney, NSW 2006, Australia}

\author{Jin Min Yang}
\email[Electronic address:]{jmyang@itp.ac.cn}
\affiliation{State Key Laboratory of Theoretical Physics, Institute of Theoretical Physics, Academia Sinica, Beijing 100190, China}

\date{\today}

\begin{abstract}
Light top squarks (stops) and light higgsinos are the key features of natural SUSY, where the higgsinos $\tilde{\chi}^{\pm}_{1}$ and $\tilde{\chi}^0_{1,2}$ are nearly degenerate and act as the missing transverse energy ($\slashed E_T$) at the LHC. Besides the strong production, the stop can be produced via the electroweak interaction. The determination of the electroweak properties of the stop is an essential task for the LHC and future colliders. So in this paper, we investigate the single stop ($\tilde{t}_1$) production $pp \to \tilde{t}_1+\slashed E_T$ in the natural SUSY at the LHC, which gives the monotop signature $t+\slashed E_T$ from $\tilde{t}_1\to t \tilde{\chi}^0_{1,2}$ or the monobottom signature $b+\slashed E_T$ from $\tilde{t}_1 \to b \tilde{\chi}^+_{1}$. We perform Monte Carlo simulations for these signatures and obtain the results:
(1) The signal $b+\slashed E_T$ has a better sensitivity than $t+\slashed E_T$ for probing natural SUSY;
(2) The parameter region with a higgsino mass 100 GeV$\lesssim \mu \lesssim$ 225 GeV and stop mass
$m_{\tilde{t}_1} \lesssim$ 620 GeV, can be probed through the single stop production with $S/\sqrt{B} > 3$ and $4\% \lesssim S/B \lesssim19\%$ at 14 TeV HL-LHC with an integrated luminosity of 3000 fb$^{-1}$.

\end{abstract}
\pacs{Valid PACS appear here}
\maketitle


\section{Introduction}
The search for supersymmetry (SUSY) is a long-standing important task in particle physics.
One prime motivation for weak-scale SUSY is that it protects the Higgs vacuum expectation
value without unnatural fine-tuning of the theory parameters.
In the minimal supersymmetric standard model (MSSM), only a small subset of the supersymmetric partners
strongly relates with the naturalness of the Higgs potential \cite{bg}.
This can be seen from the minimization of the Higgs potential \cite{mz}:
\begin{eqnarray}
\frac{M^2_{Z}}{2}&=&\frac{(m^2_{H_d}+\Sigma_{d})-(m^2_{H_u}+
\Sigma_{u})\tan^{2}\beta}{\tan^{2}\beta-1}-\mu^{2} \nonumber \\
&\simeq&-\mu^{2}-(m^2_{H_u}+\Sigma_{u}),
\label{minimization}
\end{eqnarray}
where $\mu$ is the higgsino mass parameter, and $m^2_{H_d}$ and $m^2_{H_u}$ denote the weak scale
soft SUSY breaking masses of the Higgs fields. A moderate or large $\tan\beta\equiv v_u/v_d$
is assumed in the last approximate equality. $\Sigma_{u}$ and $\Sigma_{d}$ arise from the radiative
corrections to the Higgs potential, and the one-loop dominant contribution to $\Sigma_{u}$ is given
by \cite{tata}
\begin{eqnarray}
\Sigma_u \sim \frac{3Y_t^2}{16\pi^2}\times m^{2}_{\tilde{t}_i}
\left( \log\frac{m^{2}_{\tilde{t}_i}}{Q^2}-1\right)\;.
\label{rad-corr}
\end{eqnarray}
In order to obtain the observed value of $M_Z$ without large cancelations in Eq.~(\ref{minimization}),
each term on the right hand side should be comparable in magnitude.
Thus, the higgsino mass $\mu$ must be of the order of $\sim 100-200$ GeV and
the requirement of $\Sigma_u \sim M_Z^2/2$ produces an upper bound on the stop mass, which is about
500 GeV \cite{stop-mass,jm} (a 125 GeV Higgs mass can be achieved by a large stop trilinear coupling without very heavy stops in the MSSM or achieved by extending the MSSM with additional D-terms or F-terms \cite{langacker,mohapatra,cao-mssm-nmssm}).
In addition, since the gluino contributes to $m_{H_u}$ at two-loop level, it is also upper bounded by the
naturalness \cite{gluino-mass} (however, the direct LHC searches have pushed the gluino up to TeV
scale \cite{gluino-lhc,cheng} while the recent ATLAS Z-peaked excess may indicate a gluino around
800 GeV \cite{z-peak}).

So to test SUSY naturalness, the crucial task is to search for light stops or  higgsinos.
The search strategy for the pair productions of these nearly degenerate higgsinos has been recently
studied \cite{giudice,nath,martin,higgsino-world-0,higgsino-world-0.5,higgsino-world-1,higgsino-world-1.5,higgsino-world-2,higgsino-world-3,higgsino-world-4,higgsino-world-5,higgsino-world-6,higgsino-world-7,higgsino-world-8}.
During the LHC run-1, the ATLAS and CMS collaborations have performed the extensive searches for the stops through the gluino-mediated stop production \cite{atlas-gluino-stop,cms-gluino-stop}
or the direct stop pair production \cite{atlas-stop,cms-stop}.
Meanwhile, many theoretical studies that aim for improving the LHC sensitivity to a light stop have been proposed \cite{th-stop-1,th-stop-2,th-stop-3,th-stop-4,th-stop-5,th-stop-6,th-stop-7,th-stop-8,th-stop-9,th-stop-10,th-stop-11,th-stop-12,th-stop-13,th-stop-14,th-stop-15,th-stop-16}.
The current LHC constraints indicate a stop mass bound of hundreds of GeV \cite{stop-mass-0,stop-mass-1,stop-mass-2,stop-mass-3,stop-mass-4,stop-mass-5,stop-mass-6,stop-mass-7,stop-mass-8,stop-mass-9,stop-mass-10}, however, those results are affected by the stop polarization states and branching ratios. The constraints on the right-handed stop from the LHC run-1 direct searches \cite{atlas-stop,cms-stop} are usually weakened by the branching ratio suppression, which can still be as light as 230 GeV in some compressed region \cite{rstop}. If the stop mass is heavier than about 450 GeV, it is allowed in most parameter space. So in our work, we require the stop mass be heavier than 450 GeV \cite{rstop}.

Usually, the stop pair production provides the most sensitive way to search for the stop at the LHC. However, the stop can also participate in the electroweak interaction processes. The determination of the electroweak properties of the stop is an essential task for the LHC and future colliders. In this work, we study the single stop production $pp \to \tilde{t}_1+\slashed E_T$ in the natural SUSY at the LHC. The observation of the single stop production will test the electroweak properties of the stop and the naturalness of the supersymmetry. In the following, we will perform the Monte Carlo simulations for the single stop production and examine its sensitivity at the LHC.

\begin{figure}[ht]
\centering
\includegraphics[width=8.0cm]{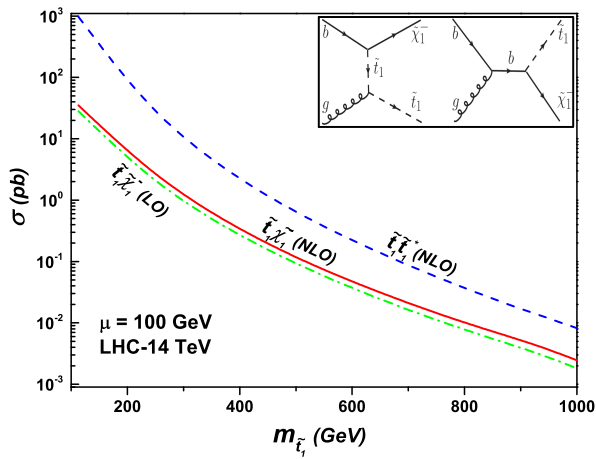}
\vspace{-0.3cm}
\caption{The cross sections of $\tilde{t}_1\tilde{t}^*_1$ and $\tilde{t}_1\tilde{\chi}^-_1$ productions
at the 14 TeV LHC for $\tan\beta=50$ and degenerate higgsinos with mass $\mu = 100$ GeV. The contribution
of conjugate process $\tilde{t}^*_1\tilde{\chi}^+_1$ is included.}
\label{cx}
\end{figure}

\section{Calculations and simulations}
At the LHC, the single stop production is induced by the electroweak interaction and proceeds through the
following process (see Fig.~\ref{cx} for the corresponding Feynman diagrams):
\begin{eqnarray}
pp \to \tilde{t}_1 \tilde{\chi}^{-}_{1}.
\end{eqnarray}
Since in natural SUSY the light higgsinos are nearly degenerate, the decay products of
$\tilde{\chi}^-_1 \to W^* \chi^{0}_{1}$ will carry small energies and, hence, are too soft
to be observed in the detector. Thus, the associated production of $\tilde{t}_1\tilde{\chi}^-_1$
can be identified as $\tilde{t}_1+\slashed E_T$, which provides a distinctive signature at the LHC.

In Fig.~\ref{cx}, we show the next-to-leading order (NLO) cross sections of the stop pair
and the single stop productions for $\mu=100$ GeV at 14 TeV LHC by using the packages \textsf{Prospino2} \cite{prospino}
and \textsf{MadGolem} \cite{madgolem}, respectively. The renormalization and factorization scales are taken as
the half average of the final states masses. In the calculations of $\tilde{t}_1\tilde{\chi}^-_1$, we use the
LO and NLO parton densities given by CTEQ6L1 and CTEQ6M with five active flavors \cite{cteq6}. The contribution
of the conjugate process $\tilde{t}^*_1\tilde{\chi}^+_1$ is included. Except for the higgsino mass parameter
$\mu$ and right-handed stop soft mass $m_{U_3}$, we assume other soft supersymmetric masses at 1 TeV, and use the
packages \textsf{SOFTSUSY-3.3.9} \cite{softsusy} and \textsf{MSSMCalc} \cite{mssmcalc} to calculate masses,
couplings and branching ratios of the sparticles. Since the cross section of single stop
production is not sensitive to $\tan\beta$, we take $\tan\beta=50$ for simplicity.
We find that the single stop cross section can still reach about 200 fb when $m_{\tilde{t}_1} \simeq 450$ GeV.
The NLO $K$-factor of the process $pp \to \tilde{t}_1\tilde{\chi}^-_1$ ranges from 1.25 to 1.33.
When the stop becomes heavy, the single stop production cross section will decrease, but slower than
the pair production, due to the kinematics.

Next, we investigate the LHC observability of the single stop signatures with the sequent decays
$\tilde{t}_{1} \to t \tilde{\chi}^{0}_{1,2}$ and $\tilde{t}_{1} \to b \tilde{\chi}^{+}_{1}$:
\begin{eqnarray}
&&pp \to \tilde{t}_1 \tilde{\chi}^{-}_{1} \to t \tilde{\chi}^0_{1,2} \tilde{\chi}^-_1 \to bjj+ \slashed E_T,\label{monotop}\\
&&pp \to \tilde{t}_1 \tilde{\chi}^{-}_{1} \to b \tilde{\chi}^+_{1} \tilde{\chi}^-_1 \to b+ \slashed E_T.\label{monob}
\end{eqnarray}
For the decay $\tilde{t}_1 \to t \tilde{\chi}^0_{1,2}$, the SM backgrounds to the signal $bjj+ \slashed E_T$ are from the semi- and full-hadronic $t\bar{t}$ events \cite{fuks-1,fuks-2,chongsheng}\footnote{Hadronic monotop has received special attention since its signature offers the possibility to use top reconstruction as a tool to reject the backgrounds. In contrast, the leptonic monotops is believed to be more challenging since the branching fraction of the leptonic top quark decay is smaller and since there are two different sources of missing transverse energy, namely a neutrino coming from the top quark decay and the new invisible state. In Ref.\cite{fuks-2}, the authors comparatively studies these two channels and found that the sensitivity of both channels are very similar.}, where the undetected lepton and the limited jet energy resolution will lead to the relatively large missing transverse energy. The processes $W+{\rm jets}$ and $Z+{\rm jets}$ can fake the signal when one of those light-flavor jets are mis-tagged as a $b$-jet. Also, the single top can mimic our signal when the lepton from the $W$ boson decay is missed at the detector. While $t \bar{t}+V$ backgrounds are not considered in our simulations due to their small missing energy or cross sections compared to the above backgrounds.

\begin{figure}[ht]
\centering
\includegraphics[width=9.0cm]{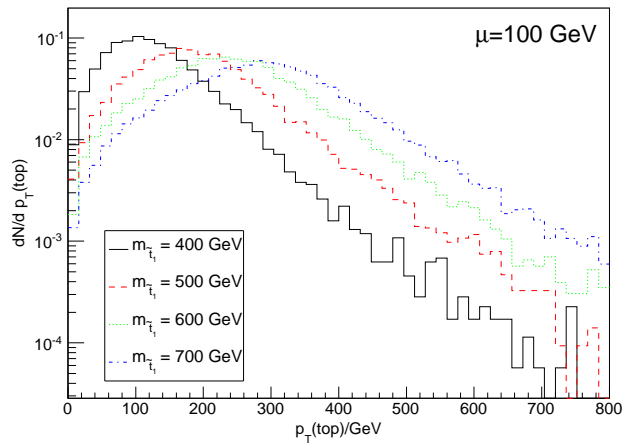}
\vspace{-0.7cm}
\caption{The parton-level $p_T$ distribution of the top quark in the channel $\tilde{t}_1 \to t \tilde{\chi}^0_{1,2}$
for $\mu=100$ GeV at 14 TeV LHC.}
\label{top-pt}
\end{figure}
In Fig.~\ref{top-pt}, we present the parton-level $p_T$ distribution of the top quark in the channel
$\tilde{t}_1 \to t \tilde{\chi}^0_{1,2}$ for $\mu=100$ GeV at 14 TeV LHC. It can be seen that, with the increase
of stop mass, the top quark produced from stop decay is boosted and has larger $p_T$. So, in the analysis of
$\tilde{t}_1 \to t \tilde{\chi}^0_{1,2}$ channel, we respectively adopt \textsf{HEPTopTagger} \cite{th-stop-2}
and normal hadronic top reconstruction methods for each sample to identify the top quark in the final states
and present our results with the best one. The detailed analysis strategies are the followings:
\begin{itemize}

\item Events with any isolated leptons are rejected;

\item \textbf{Method-1}: We use Cambridge-Aachen (CA) algorithms~ \cite{ca} in \textsf{Fastjet}~\cite{fastjet} to cluster the jets with $R=1.5$ to obtain the \textit{top-jet} candidates. Each candidate must have the top quark substructure required by the \textsf{HEPTopTagger}. The $b$-tagging is also imposed in the \textit{top-jet} reconstruction. Other energy deposits outside the \textit{top-jet} are further reconstructed as the normal jets by using anti-$k_t$ algorithm with $R=0.4$ \cite{anti-kt}. The top window used in our analysis is $150< m_t <200$ GeV. While the $W$ window is taken as the default value in \textsf{HepTopTagger};
\item \textbf{Method-2}: In normal hadronic top quark reconstruction, a pair of jets is selected with the invariant mass $m_{jj} > 60$ GeV and the smallest $\Delta R$. A third jet closest to this di-jet system is used to constitute the top quark candidate. Among these three jets, at least one $b$-jet and $\Delta\phi(\slashed E_T, p_T(b_1)) > 1$ is required. The anti-$k_t$ algorithm is used for jet clustering with $R=0.4$;

\item We keep the events with the exact one reconstructed top quark and require 150 GeV $< m^{\rm rec}_t < $ 200 GeV;

\item The extra leading jet $j_1$ outside the reconstructed top quark object is vetoed if $p_T(j_1) > 30$ GeV and $|\eta(j_1)|<2.5$;

\item We define the signal regions according to $(\slashed E_T, p_T(j_{\rm top}))$ cuts: (200, 100), (250, 150), (300, 200), (350, 250) for Method-1, and $(p_T(b), \slashed E_T)$ cuts: (200, 50), (250, 50), (300, 100), (350, 100) GeV for Method-2.

\end{itemize}

For the decay $\tilde{t}_1 \to b \tilde{\chi}^+_{1}$, the SM backgrounds to the signal $b+ \slashed E_T$
are dominated by the processes $W+{\rm jets}$ and $Z+{\rm jets}$ when the light-flavor jets are mis-identified as $b$-jets \cite{liantao}.
The $t\bar{t}$ events become the sub-leading backgrounds due to their large multiplicity. The signal events are
selected to satisfy the following criteria:
\begin{itemize}
\item Events with any isolated leptons are rejected;
\item We require exact one hard $b$-jet in the final states, but allow an additional softer jet with $p_T(j_1)<30$ GeV
and $\Delta\phi(\slashed E_T, p_T(j_1)) > 2$. Since the hardness of $b$-jet from stop decay depends on the mass splitting
between $\tilde{t}_1$ and $\tilde{\chi}^-_1$, we define three signal regions for each sample according to
$(\slashed E_T, p_T(b))$ cuts: (100, 70), (150, 100) and (250, 200) GeV.
\end{itemize}
Finally, we use the most sensitive signal region (with the highest $S/\sqrt{B}$)
for each decay mode and show our results in Fig.~\ref{ss-top} and Fig.~\ref{ss-bottom}, respectively.
In our study, we omitted the QCD multijet backgrounds,
whose correct treatment needs the experimental data-driven methods and hence depends on the
realistic detector environments of the 14 TeV LHC.
As discussed in  \cite{fuks-2,liantao,cms-monotop},
the requirements of high $p_T(b_1)$ and large $\slashed E_T$ with a seperation
$\Delta\phi(\slashed E_T, p_T(j_1))$ can usually be expected to greatly reduce the fake contamination
from the QCD backgrounds and allow for a good signal selection efficiency.

The parton level signal and background events are generated with \textsf{MadGraph5} \cite{mad5}, where $W/Z+{\rm jets}$ is matched up to 3 jets by using MLM matching scheme \cite{mlm} and setting $xqcut=30$ GeV. For the value of $qcut$ in matching, we take it to $max(xqcut+5, xqcut*1.2)$ \cite{matching} in our simulation. We perform parton shower and fast detector simulations with \textsf{PYTHIA} \cite{pythia} and \textsf{Delphes} \cite{delphes}. We assume the $b$-jet tagging efficiency as 70\% \cite{cms-b} and a misidentification efficiency of $c$-jets and light jets as 10\% and 0.1\%, respectively. The cross section of $t\bar{t}$ is normalized to the approximately next-to-next-to-leading order value $\sigma_{t\bar{t}}=920$ pb \cite{ttbar}.

\section{Results and discussions}
\begin{table}[ht]
\begin{center}
\caption{The cross sections of $V+{\rm jets}$, $t\bar{t}$ and $\tilde{t}_1(\to t \tilde{\chi}^0_{1,2})\tilde{\chi}^-_1$
for a benchmark point $(m_{\tilde{t}_{1}}, \mu)=(611,100)$ GeV and $\tan\beta=10$ in Method-1 and Method-2 at 14 TeV
LHC with ${\cal L}=3000$ fb$^{-1}$. The cross sections are in unit of fb.}
\begin{tabular}{|c|c|c|c|c|c|c|c|}
\hline
 cuts &$W+{\rm jets}$ &$Z+{\rm jets}$ &$t\bar{t}$ &$tW$ & $S$ &$S/B$ &$S/\sqrt{B}$ \\
\hline
Method-1 &$<10^{-2}$ &0.29  &2.20  &0.80 &0.13 &4.0\% &$3.9$ \\
\hline
Method-2 &$<10^{-2}$ &0.59 &0.55 &0.24 &0.044 &3.2\% &$2.1$ \\
\hline
\end{tabular}\label{tab-top}
\end{center}
\end{table}
In Table \ref{tab-top}, we compare the results of $pp \to \tilde{t}_1(\to t \tilde{\chi}^0_{1,2})\tilde{\chi}^-_1$
for a benchmark point $(m_{\tilde{t}_{1}}, \mu)=(611,100)$ GeV in Method-1 and Method-2 at 14 TeV LHC.
From this table we can see that the $Z+{\rm jets}$ background in Method-1 is smaller than in Method-2, while the
$t\bar{t}$ background in Method-1 is larger than in Method-2. However, the signal events can be more kept in
Method-1 than in Method-2. So the overall effects make the Method-1 have a better sensitivity in reconstructing
the top quark in the region with large mass splitting between $\tilde{t}_1$ and $\tilde{\chi}^-_1$. At 14 TeV LHC
with ${\cal L}=3000$ fb$^{-1}$, the statistical significance $S/\sqrt{B}$ for our benchmark point can reach
$3.9\sigma$ ($2.1\sigma$) with $S/B=4.0\%$ ($3.2\%$) in Method-1(2).

\begin{table}[ht]
\begin{center}
\caption{The cross sections of $V+{\rm jets}$, $t\bar{t}$ and $\tilde{t}_1(\to b \tilde{\chi}^+_1)\tilde{\chi}^-_1$ for
a benchmark point $(m_{\tilde{t}_{1}}, \mu)=(496, 200)$ GeV and $\tan\beta=10$ at 14 TeV LHC with ${\cal L}=3000$ fb$^{-1}$.
The cross sections are in unit of fb.}
\begin{tabular}{|c|c|c|c|c|c|}
\hline
$W+{\rm jets}$ &$Z+{\rm jets}$ &$t\bar{t}$ & $S$ &$S/B$ &$S/\sqrt{B}$ \\
\hline
$<10^{-2}$ &2.77 &1.10 &0.20 &5.1\% &$5.5$\\
\hline
\end{tabular}\label{tab-sbottom}
\end{center}
\end{table}
In Table \ref{tab-sbottom}, we show the cross sections of $V+{\rm jets}$, $t\bar{t}$ and
$\tilde{t}_1(\to b \tilde{\chi}^+_1)\tilde{\chi}^-_1$ for a benchmark point $(m_{\tilde{t}_{1}}, \mu)=(496, 200)$ GeV at
the 14 TeV LHC. Different from $\tilde{t}_1 \to t \tilde{\chi}^0_{1,2}$ channel, $Z+{\rm jets}$ background is dominant over
$t\bar{t}$ since only one hard $b$-jet is required in the final state. From Table \ref{tab-sbottom} we can see that
$S/\sqrt{B}$ and $S/B$ can reach $5.5$ and $5.1\%$ for ${\cal L}=3000$ fb$^{-1}$, respectively.

\begin{figure}[ht]
\centering
\includegraphics[width=8.0cm]{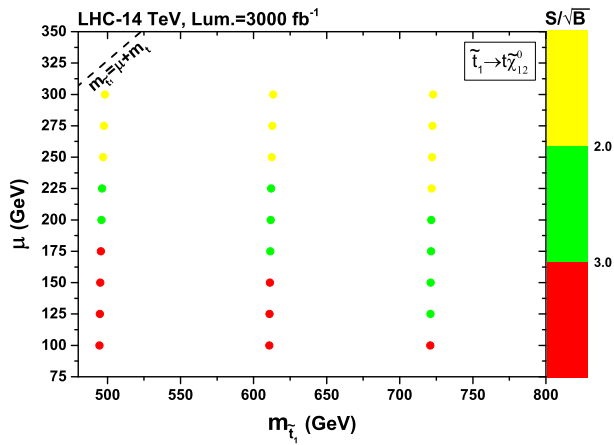}
\vspace{-0.3cm}
\caption{The dependence of the significance of the channel $\tilde{t}_1 \to t \tilde{\chi}^0_{1,2}$ on the higgsino
mass $\mu$ and stop mass $m_{\tilde{t}_1}$ at the 14 TeV LHC with $\mathcal{L} = 3000$ fb$^{-1}$.}
\label{ss-top}
\end{figure}
In Fig.~\ref{ss-top}, we display the dependence of statistical significance $S/\sqrt{B}$ of the channel
$\tilde{t}_1 \to t \tilde{\chi}^0_{1,2}$ on the higgsino mass $\mu$ and stop mass $m_{\tilde{t}_1}$ at 14 TeV LHC
with $\mathcal{L} = 3000$ fb$^{-1}$. We can see that values of $S/\sqrt{B}$ decrease with the increase of $\mu$ because
of the cut efficiency reduction. When the stop becomes heavy, the cross section of $\tilde{t}_1\tilde{\chi}^-_1$ is
suppressed. However, as a result of the application of \textsf{HEPTopTagger} method, more signal events can be kept,
in particular in the mass range 450 GeV $\lesssim m_{\tilde{t}_1} \lesssim $ 650 GeV. Therefore, when $\mu \lesssim 150$ GeV,
the stop mass $m_{\tilde{t}_1} \lesssim$ 610 GeV can be probed at $\gtrsim 3\sigma$ statistical significance
with $S/B \lesssim 8\%$.

\begin{figure}[ht]
\centering
\includegraphics[width=8.0cm]{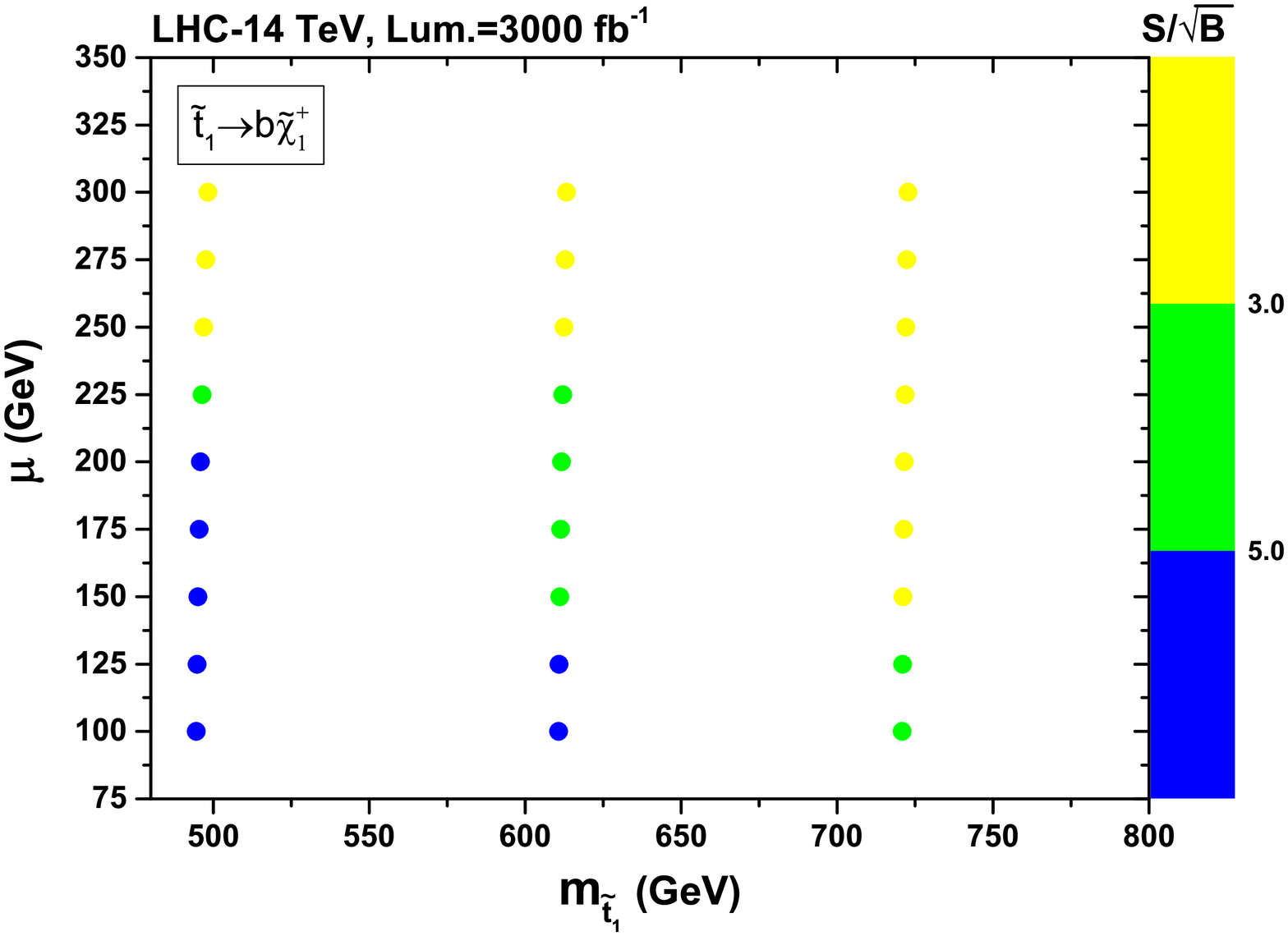}
\vspace{-.3cm}
\caption{Same as Fig.~\ref{ss-top}, but for the decay channel $\tilde{t}_1 \to b \tilde{\chi}^+_{1}$.}
\label{ss-bottom}
\end{figure}
In Fig.~\ref{ss-bottom}, the statistical significance $S/\sqrt{B}$ of the channel $\tilde{t}_1 \to b \tilde{\chi}^+_{1}$
is presented on the plane of higgsino mass $\mu$ versus stop mass $m_{\tilde{t}_1}$ at 14 TeV LHC with
$\mathcal{L} = 3000$ fb$^{-1}$. It can be seen that the sensitive stop region lies
in 450 GeV $\lesssim m_{\tilde{t}_1} \lesssim$ 620 GeV, where a hard $b$-jet ($p_T > 200$ GeV) and the sizable
$\slashed E_T$ ($\slashed E_T > 250$ GeV) can be used to effectively suppress the backgrounds. But when the stop
mass increases, $S/\sqrt{B}$ will rapidly decrease. We see that the higgsino
mass 100 GeV$\lesssim \mu \lesssim$ 225 GeV and the stop mass $ m_{\tilde{t}_1} \lesssim$ 620 GeV
can be covered at $\gtrsim 3\sigma$ statistical significance with $S/B$ varying from $4\%$ to $19\%$.

\section{Conclusions}
even its exclusion limit for the stop may be not as good as the pair production. However, the stop can participates in the strong interaction processes but also in the electroweak interaction processes, the determination of the electroweak properties of the stop is an essential task for the LHC and future colliders. In this work we propose to probe natural SUSY by using the electroweak single-stop production $pp \to \tilde{t}_1+\slashed E_T$ at the LHC (here the missing energy is from the nearly degenerate higgsinos). By analyzing the decay channels
of the stop $\tilde{t}_1 \to t \tilde{\chi}^0_{1,2}$ and $\tilde{t}_1 \to b \tilde{\chi}^+_{1}$, we obtain the
obervations:
(1) The decay $\tilde{t}_1 \to b \tilde{\chi}^+_{1}$ has a better sensitivity than $\tilde{t}_1 \to t \tilde{\chi}^0_{1,2}$ ;
(2) The parameter region with a higgsino mass 100 GeV$\lesssim \mu \lesssim$ 225 GeV and the stop mass $ m_{\tilde{t}_1} \lesssim$ 620 GeV
can be covered
with $S/\sqrt{B} > 3$ and $4\% \lesssim  S/B \lesssim19\%$ at 14 TeV HL-LHC with an integrated luminosity of 3000 fb$^{-1}$. So the searches for the single stop production will directly test the naturalness of the supersymmetry and the electroweak properties of the stop.

\section{acknowledgments}
Lei Wu thanks David Lopez-Val and Dorival Goncalves for providing us the \textsf{MadGolem}
package.
This work was partly supported by the Grant-in-Aid for Scientific Research (No.~24540246)
from Ministry of Education, Culture, Sports, Science and Technology (MEXT) of Japan,
by the Australian Research Council,
by the CAS Center for Excellence in Particle Physics (CCEPP),
by the National Natural Science Foundation of China (NNSFC) under grants
Nos. 11305049, 11275057, 11375001, 11405047, 11275245, 10821504 and 11135003,
and by Specialized Research Fund for the Doctoral Program of Higher Education
under Grant No.20134104120002.

\end{document}